# The AI Act proposal: a new right to technical interpretability?

CHIARA GALLESE*

SUMMARY: 1. Introduction. – 2. The concepts of "Technical Interpretability" and "Explainability". – 3. The transparency principle and the "Right to Explanation". – 4. Articles 13 and 14 of the AI Act Proposal. – 5. The "Right to Technical Interpretability" as a fundamental right.

## 1   *Introduction*

The debate about the concept of "right to explanation" is the subject of a wealth of literature. It has focused in the legal scholarship on art. 22 GDPR and in the technical scholarship on explaining the output of a certain model.

The purpose of this paper is to investigate if the new provisions introduced by the proposal for a Regulation laying down harmonised rules on artificial intelligence (AI Act), in combination with Convention 108+ and GDPR, are enough to indicate the existence of a "right to technical explainability" and, if not, whether the EU legal system should include it in its current legislation.

Before examining the legal aspects of "interpretability", it is necessary to provide clear definitions and to differentiate it from "explainability" (Section 2). From a legal point of view, the difference between the two concepts is blurred and only a "right to explanation" has been theorized so far (Section 3). However, the new AI Act proposal, in its original version, provided some important transparency principles that might influence the way in which AI systems are built from a technical point of view (Section 4). In the last section of this article, the existence of a "right to technical interpretability" is theorized.

## 2   *The concepts of "Technical Interpretability" and "Explainability"*

Before examining the different sources of law that contain the right of explanation, it is beneficial to consider about the distinction between *explainability* and *interpretability*. Although some researchers have attempted to provide a thorough definition of the two words[2], there is still a lot of ambiguity in the literature regarding these notions, and they are frequently used indiscriminately[3]. The two terminologies are treated as separate notions in this essay.

---

*Postdoctoral Researcher, Department of Electrical Engineering, Eindhoven University of Technology, Eindhoven, The Netherlands; Department of Mathematics and Geosciences, University of Trieste; School of Engineering, Carlo Cattaneo University - LIUC, Varese, Italy. Funded by the REMIDE project, Carlo Cattaneo University - LIUC and by the UNI 4 JUSTICE Project

[2] Maximilian A Köhl et al. "Explainability as a non-functional requirement". In: *2019 IEEE 27th International Requirements Engineering Conference (RE)*. IEEE. 2019, pp. 363– 368; Larissa Chazette and Kurt Schneider. "Explainability as a non-functional requirement: challenges and recommendations". In: *Requirements Engineering* 25.4 (2020), pp. 493–514; Larissa Chazette, Wasja Brunotte, and Timo Speith. "Exploring explainability: A definition, a model, and a knowledge catalogue". In: *2021 IEEE 29th International Requirements Engineering Conference (RE)*. IEEE. 2021, pp. 197–208.

If we consider the definition provided by Rudin (2022)[4], *explainability* occurs when "a second (post hoc) model is created to explain the first black box model". In this scenario, there are two models: the first is opaque since it is impossible to understand why it produced a particular output, and the second, which produces a human-intelligible output, is used to attempt to piece together the factors that led to the first model's output. Explainability, then, is the attempt to explain the output of a black box.

According to the author, in certain circumstances explainability may not be a reliable method, because explainable machine learning techniques offer explanations that are not accurate representations of what the original model computes, and therefore suggest an incorrect explanation. In fact, continues the author, they are unable to accurately reconstruct the original model. If the explanation accurately described what the original model computed, it would be identical to the original model, negating the very necessity for the original model in the first place: in other words, the second model would be interpretable[5].

Babic et al. (2021)[6] also present some doubts, pointing out three circumstances: first, as it is frequently the most accurate, the opaque function of the black box continues to serve as the foundation for AI/ML choices. Second, there cannot be a perfect approximation between the white box and the black box; otherwise, there would be no distinction between the two (as noted by Rudin). Additionally, fitting the black box, frequently just locally, is the main priority rather than accuracy. Finally, the justifications offered are post hoc.

This draws attention to a crucial point that should be taken into account: based on current knowledge, black box models are typically more accurate than alternative models. Nowadays, deep learning outperforms white box models in a number of fields, including imaging. But in some circumstances, interpretable models might be just as accurate as black boxes[7]. Petch et al.[8] suggest a general guideline for choosing whether to utilize black boxes:

---

[3] Cynthia Rudin et al. "Interpretable machine learning: Fundamental principles and 10 grand challenges". In: *Statistics Surveys* 16 (2022), pp. 1–85.

[4] Cynthia Rudin. "Stop explaining black box machine learning models for high stakes decisions and use interpretable models instead". In: *Nature Machine Intelligence* 1.5 (2019), pp. 206–215.

[5] Ibid.

[6] Boris Babic et al. "Beware explanations from AI in health care". In: Science 373.6552 (2021), pp. 284–286

[7] Evangelia Christodoulou et al. "A systematic review shows no performance benefit of machine learning over logistic regression for clinical prediction models". In: *Journal of clinical epidemiology* 110 (2019), pp. 12–22.

[8] Jeremy Petch, Shuang Di, and Walter Nelson. "Opening the black box: the promise and limitations of explainable machine learning in cardiology". In: *Canadian Journal of Cardiology* (2021).

> "[…] data scientists should train models using both interpretable and black-box methods to assess whether there is, in fact, an accuracy vs interpretability tradeoff in the specific case on which they are working. If there is no meaningful difference in accuracy between an interpretable model and a black box, an interpretable method should be used. However, if a black-box model does provide a higher degree of accuracy, the stakes of the decision should be considered. If the decision that will be informed by the model is a relatively low stake, a small improvement in accuracy may justify the use of a black box. However, if the stakes are high, it is reasonable to require a greater improvement in accuracy before sacrificing interpretability. Ideally, gains in accuracy from black-box methods should be sufficient to translate into meaningful improvements in clinical outcomes such as reduced morbidity or mortality. If the use of a black box model can be justified, explainability techniques should be employed to make the model and its predictions as transparent as possible, but clinicians should be aware of their limitations and be cautious of overinterpreting, which can lead to narrative fallacies".

The term *interpretability* describes a model that is understandable in a way that enables people to comprehend the substantial (not mere technical) justification that resulted in a specific output. "Justifiable" could be used as a synonym here; someone must be able to understand the logic involved in the decision, whether they be the modeler, the domain expert, the final user, or the individual who will be affected by the outcome.

It is important to note that a model's mere mathematical justification and comprehension of the model structure (such as being able to look at the source code) are insufficient for humans to comprehend why the model came to a particular conclusion. Knowing the source code of a neural network, the number of parameters (weights), layers, or features, without knowing the reason why a relationship between data is found, may not enable humans to understand the output.

Consider a classification algorithm that categorizes upcoming patients of a particular condition into healthy or unhealthy based on a data set comprising millions of physiological characteristics of past patients. In order to determine whether the system classifies a new patient as healthy or ill, doctors may enter data from the patient, including blood levels, symptoms, anamnesis, genomic information, lifestyle choices, age, number of children, ethnicity, weight, height, number of sleep hours, job, place of birth, etc. However, because of the numerous and intricate features, parameters, and layers that are employed in producing the output, the system is unable to determine the cause of the patient's illness, such as the fact that the blood levels are abnormal for someone of the patient's age, ethnicity, weight, and daily amount of exercise. It's even possible that the system depended on unnecessary features by coincidence, such as the ID number or the date of admission to the hospital. Nor the physician nor the modeller has any mean to determine this. Das and Rad[9] consider that:

> "The large number of parameters in Deep Neural Networks (DNNs) make them complex to understand and undeniably harder to interpret. Regardless of the cross-validation accuracy or other evaluation parameters which might indicate a good learning performance, deep learning (DL) models could inherently learn or fail to learn representations from the data which a human might consider important. [...] Hence, often the ability to interpret AI decisions are deemed secondary in the race to achieve state-of-the-art results or crossing human-level accuracy".

In addition, Rudin[10] notes that interpretability is not a black and white concept, but rather a spectrum:

> "There is a spectrum between fully transparent models (where we understand how all the variables are jointly related to each other) and models that are lightly constrained in model form (such as models that are forced to increase as one of the variables increases, or models that, all else being equal, prefer variables that domain experts have identified as important[...])".

There are differences in the literature about *what* should be explainable and in what context. Chazette et al.[11] summarize the elements that, according to the literature and their own analysis, should be explained, such as the inference processes for certain problems, the relationships between the inputs and outputs, parameters and data structures, intentions, behaviors in real-world, underlying criteria for the decision, predictive accuracy, and user preferences.

Often, what is meant with the term "Exaplainable Artificial Intelligence (XAI)" is instead "Intelligible AI". The concept of "intelligibility" is crucial because it encompasses a wide range of considerations that need to be made, including cultural differences[12], mental capacity, age, educational attainment, experience and expertise, preferences in visualization and design, and numerous other variables that may affect a recipient's capacity to comprehend a given output. As stated in Article 13 of the GDPR, it is essential to adapt to the addressee (the audience), and this is especially true when there is an impact on their life. Numerous scholars have emphasized the significance of the addressee's comprehension, which might vary depending on the circumstances[13]. However, in this article we will only focus on the inherent technical intelligibility of decisions, not on all the other elements that make an output intelligible, as important.

---

[9] Arun Das and Paul Rad. "Opportunities and challenges in explainable artificial intelligence (xai): A survey". In: *arXiv preprint arXiv:2006.11371* (2020).

[10] Rudin, "Stop explaining black box machine learning models for high stakes decisions and use interpretable models instead".

[11] Chazette, Brunotte, and Speith, "Exploring explainability: A definition, a model, and a knowledge catalogue".

[12] Including language: different languages may have different ways of expressing a concept. Localization is an important element of the transparency principle.

For instance, considering a system that predicts the likelihood of not being able to pay back a mortgage and is used by a financial institution to deny credit, we would consider it interpretable only if it made clear which financially significant factors—such as wage, job type, age, concurrent loans, marital status, and education—were used by the model to produce the output, what relationship were found between them (e.g., educated persons are more likely to have high incomes), and which ones were given a higher weight than others (for example, the system could weight the past mobility as an unfavourable condition and weight it more that an advanced age). Even if this explanation were to be given in mathematical terms, the modellers would still be able to translate it such that bank employees could understand it, and the personnel would then be able to explain it to the mortgage applicant in plain language.

3   *The transparency principle and the "Right to Explanation"*

Transparency is a key principle and an overarching obligation in the whole EU legislation and in particular within the Digital Strategy, but it is also an important ethical and legal requirement provided by national laws and guidelines in some fields relating to high-risk systems.

The "right of explanation" in GDPR is part of transparency: data subjects have the right to receive information about the rationale behind or the criteria relied on in reaching an automated decision that has an impact on their life, and about the significance and envisaged consequences of the processing of their data, as provided by Articles 13 and 14 of GDPR. Controllers must provide meaningful information about the logic involved in the decision process, not necessarily a complex explanation of the algorithms used or the disclosure of the full source code[14], but a "sufficiently comprehensive explanation that allows the data subject to understand the reasons for the decision"[15]. This concept is closer to interpretability than to explainability: it is more important that data subjects can understand what a model did than how it did it from a technical point of view.

---

[13] Marco Tulio Ribeiro, Sameer Singh, and Carlos Guestrin. """ Why should i trust you?" Explaining the predictions of any classifier". In: *Proceedings of the 22nd ACM SIGKDD international conference on knowledge discovery and data mining*. 2016, pp. 1135–1144; Diogo V Carvalho, Eduardo M Pereira, and Jaime S Cardoso. "Machine learning interpretability: A survey on methods and metrics". In: *Electronics* 8.8 (2019), p. 832; Tim Miller. "Explanation in artificial intelligence: Insights from the social sciences". In: *Artificial intelligence* 267 (2019), pp. 1–38; Avi Rosenfeld and Ariella Richardson. "Explainability in human–agent systems". In: *Autonomous Agents and Multi-Agent Systems* 33.6 (2019), pp. 673–705; Alejandro Barredo Arrieta et al. "Explainable Artificial Intelligence (XAI): Concepts, taxonomies, opportunities and challenges toward responsible AI". in: *Information fusion* 58 (2020), pp. 82– 115; Chazette, Brunotte, and Speith, "Exploring explainability: A definition, a model, and a knowledge catalogue".

[14] Even consumer law does not require full disclosure of all the algorithms involved. Directive (EU) 2019/2161 requires transparency only regarding the main parameters used by the model.

[15] Article 29 Data protection Working Party. "Guidelines on Automated individual decisionmaking and Profiling for the purposes of Regulation 2016/679". In: *WP215* 1 (2017).

The right of explanation is also present in the Council of Europe's Convention 108+, with a broader application than in GDPR, as explained in the Explanatory Report, in Article 10:

> "Data subjects should be entitled to know the reasoning underlying the processing of their data, including the consequences of such reasoning, which led to any resulting conclusions, in particular in cases involving the use of algorithms for automated decision making including profiling. For instance, in the case of credit scoring, they should be entitled to know the logic underpinning the processing of their data and resulting in a 'yes' or 'no' decision, and not simply information on the decision itself. Without an understanding of these elements, there could be no effective exercise of other essential safeguards such as the right to object and the right to complain to a competent authority".

Transparency means that Controllers must provide to data subjects (e.g., patients) "relevant information related to fair processing, communicate and facilitate the exercise of their rights, enabling them to understand, and if necessary, challenge the data processing"[16]. However, the current legislation does not dictate the provision to data subject of a "technically faithful explanation", since not all models allow the modeller or the user to know the true reasoning behind the output. A legally compliant explanation could also be that a job applicant was rejected "because the CV does not match the minimum requirements listed in the job posting in terms of experience and education", while the truth is that the true reason is not known, or that it is only partially known but it is full of hidden biases[17]. Since the reasons are not correctly communicated to the job applicants, they would have no knowledge of the biases and could not file a discrimination lawsuit.

It must be noted that GDPR and national law also provides that the data subjects must express their informed consent. How is possible to truly and freely express consent if it is not known why the system has given a certain output? As we have seen above, only interpretability can guarantee that the substantial reasons on which the decision has been relied are known.

---

[16] Article 29 Data protection Working Party. "Guidelines on Transparency under Regulation 2016/679". In: *WP260 rev* 1 (2018).

[17] For a broader analysis on hiring software discrimination see Kelly-Lyth, Aislinn. "Challenging biased hiring algorithms". In: *Oxford Journal of Legal Studies* 41.4 (2021): 899-928.

## 4    *Articles 13 and 14 of the AI Act Proposal*

In April 2021, the European Commission published the AI Act proposal, since the specific characteristics of certain AI systems may have an impact on user safety and fundamental rights, creating new risks which need to be addressed. The most important innovation of the proposal is the establishment of four risks categories for AI systems, in order to protect citizens' fundamental rights and freedoms[18].

The risk categories are related to the degree (intensity and scope) of risk for the safety or fundamental rights of citizens. Taking inspiration from the product safety legislation, the classification of risks is based on the intended purpose and modalities for which the AI system is used, not only on their specific function.

According to the new legal framework, some AI systems are considered as 'high-risk' - in particular, AI decision support systems having an impact on important personal interests, e.g., in the case of healthcare - and must, therefore, fulfil new requirements before being put into the market or into service, including a risk management system, appropriate data governance measures and a quality management system, the use of high-quality datasets, the establishment of relevant documentation to enhance traceability, the sharing of adequate information with the user, the design and implementation of appropriate human oversight measures, and the achievement of the highest standards in terms of robustness, safety, cybersecurity, and accuracy, as well as the respect of applicable laws and regulations (e.g., GDPR).

Article 13[19] use the phrase "enable users to interpret the system's output" and mention the concept of transparency twice. We need to understand what the legislator intended to achieve through this wording. Analizying the heading of article 13, we find the words "Transparency" and "information", liked by the conjunction "and", a sign that the two concepts have separate meanings. Therefore, the law considers transparency as being a different concept than merely providing information to users, such as in the readme text file or in the technical instructions to users. However, this is not enough to say that a right to technical interpretability is enshrined in the AI Act.

In order to analyze the meaning of Article 13, we also need to have a look at the related recital. However, Recital 47 is not very helpful in providing information about Article 13, since it only adds a few specifications to its text: "To address the opacity that may make certain AI systems incomprehensible to or too complex for natural persons, a certain degree of transparency should be required for high-risk AI systems. Users should be able to interpret the system output and use it appropriately. High-risk AI systems should therefore be accompanied by relevant documentation and instructions of use and include concise and clear information, including in relation to possible risks to fundamental rights and discrimination, where appropriate". In addition, it seems to link the concept of interpretability to the mere provision of documents and instruction.

---

[18] The explanatory memorandum attached to the proposal, in fact, notes that "The use of AI with its specific characteristics (e.g. opacity, complexity, dependency on data, autonomous behaviour) can adversely affect a number of fundamental rights enshrined in the EU Charter of Fundamental Rights […] In case infringements of fundamental rights still happen, effective redress for affected persons will be made possible by ensuring transparency and traceability of the AI systems coupled with strong ex post controls".

[19] "Transparency and provision of information to users 1. High-risk AI systems shall be designed and developed in such a way to ensure that their operation is sufficiently transparent to enable users to interpret the system's output and use it appropriately. An appropriate type and degree of transparency shall be ensured […] 3. The information referred to in paragraph 2 shall specify: […] (d) the human oversight measures referred to in Article 14, including the technical measures put in place to facilitate the interpretation of the outputs of AI systems by the users […]".

Article 14 mentions the concept of interpretability when referring to the human oversight measures, prescribing that one of the measures to achieve it is to enable the user to "correctly interpret the high-risk AI system's output, taking into account in particular the characteristics of the system and the interpretation tools and methods available". Interpretability is then a mandatory, yet alternative, measure to make sure that a human is always kept in the loop to oversee the behavior of the AI system. Although this provision is not, alone, sufficient to affirm that a right to technical interpretability exists in the AI Act, it is certainly a strong argument in its favor.

Other recitals may contain relevant information regarding the degree of transparency required for high-risk systems: for example, Recital 39 states that "[…] The accuracy, non-discriminatory nature and transparency of the AI systems used in those contexts are therefore particularly important to guarantee the respect of the fundamental rights of the affected persons, notably their rights to free movement, non-discrimination, protection of private life and personal data, international protection and good administration". One might argue that without technical interpretability, or at least a very accurate explainability technique, it is impossible to guarantee the absence of discriminatory outputs and provide remedies against them.

The AI Act, overall, does not respond to our question regarding the existence of a "right to technical interpretability". The matter should be looked under the lens of a systematic interpretation, taking into consideration the sources of law that protect fundamental human rights.

5   *The "Right to Technical Interpretability" as a fundamental right*

Many academics have argued that black box algorithms shouldn't be used as normal practice in industries like medicine because of their internal opacity. This is because they can't ensure key aspects of good medical treatment[20]. Das et al.[21] believe that, due to the significant impact of data bias, trustworthiness, and adversarial examples in machine learning, it is currently not recommended to blindly accept the output of a highly predictive classifier.

---

[20] Cynthia Rudin, "Stop explaining black box machine learning models for high stakes decisions and use interpretable models instead"; Shinjini Kundu. "AI in medicine must be explainable". In: *Nature Medicine* 27.8 (2021), pp. 1328–1328.

[21] Das and Rad, "Opportunities and challenges in explainable artificial intelligence (xai): A survey".

Regarding post-hoc explainability, although it may be a useful tool, the limitations must be taken into account, as explained by Rudin[22] and Babic[23]. Vale et al. illustrate the limitations of post-hoc explainability techniques in proving discrimination, arguing that the tendency towards showing result parity that is necessary for EU non-discrimination law is absent from post-hoc explainability methodologies[24]. They argue that, because of their technical flaws, they are occasionally unstable and exhibit low fidelity, being unable to convincingly prove that there is no discrimination: the limited bias types discovered by post-hoc explainability approaches require their use to be contextualized and limited. According to them, the use of post-hoc explainability approaches is beneficial, particularly in the creation and development of models, but they might not be appropriate for use in regulatory or legal applications; as a result, they cannot be promoted as panaceas and cannot be valued solely in a vacuum without regard to more comprehensive fairness metrics. The authors believe that the substantive legal weight that post-hoc explainability procedures might be able to provide is questionable if they cannot establish *prima facie* discrimination. Therefore, they reach the same conclusion as Rudin and Babic: if a black-box model's insights and/or internal workings cannot be relied upon, it should not be used in situations where its judgments could have significant and/or long-lasting implications.

Another factor that encourages the use of interpretable models in high-risk systems, and more broadly in applications that may have a significant impact on citizens' life, is the fact that these systems frequently affect fundamental human rights, which are safeguarded by both international legal instruments and most Constitutions. This suggests that even a small chance of discrimination resulting from unintentional bias is not accepted by the legal system. Black boxes make it impossible to regulate the model output and to examine the reasoning process to see whether it is based on unfair or irrelevant criteria. Additionally, it would be exceedingly challenging to determine whether the prejudiced output reflects societal biases or the modeler's own unconscious biases or opinions, which could cause issues with determining responsibility in some legal systems (e.g., for gross negligence or willful misconduct in creating a biased model)[25].

---

[22] Cynthia Rudin, "Stop explaining black box machine learning models for high stakes decisions and use interpretable models instead".

[23] Boris Babic et al. "Beware explanations from AI in health care". In: *Science* 373.6552 (2021), pp. 284–286.

[24] Daniel Vale, Ali El-Sharif, and Muhammed Ali. "Explainable artificial intelligence (XAI) post-hoc explainability methods: Risks and limitations in non-discrimination law". In: *AI and Ethics* (2022), pp. 1–12.

[25] It is worth noting that the European Commission has recently published a proposal for a directive on AI Liability, which covers the civil liability connected to high-risk systems. The topic is, however, to broad to be addressed here. For more insights on AI liability, see Chiara Gallese, "Suggestions for a Revision of the European Smart Robot Liability Regime", Proceedings of the 4th European Conference on the Impact of Artificial Intelligence and Robotics (ECIAIR), 4.1 (2022), pp. 29-35.

Although neither GDPR nor the AI Act unambiguously dictates the use of interpretable techniques, and some authors have even been challenging the very existence of a "right to explanation"[26], in many AI applications the only way to protect citizen's fundamental rights and freedom is to employ interpretable models.

Taking into account the systematic interpretation of the EU and international legal framework surrounding high-risk systems, it is possible to argue that interpretability should be used as a standard in those fields (and even in other sensitive fields) and that black boxes should only be used in situations where it is possible to make a decision by evaluating factors other than the AI output. The very possibility of expressing informed consent and challenging the decision made on the basis of an automated decision-making system is excluded by the opacity and complexity of black boxes. The lack of technical interpretability prevents the exercise of many fundamental rights, such as the right to a fair trial, to self-determination, to non-discrimination, and more.

A "right to technical interpretability" should, therefore, be theorized at the European level, being considered a fundamental right, and embedded in the AI Act proposal.

---

[26] Sandra Wachter, Brent Mittelstadt, and Luciano Floridi. "Why a right to explanation of automated decision-making does not exist in the general data protection regulation". In: *International Data Privacy Law* 7.2 (2017), pp. 76–99.